\documentclass[conference,10pt]{IEEEtran}
\IEEEoverridecommandlockouts
\usepackage{amsmath}
\usepackage{iftex}
\usepackage[T1]{fontenc}
\ifPDFTeX
  \usepackage[utf8]{inputenc}
\fi
\usepackage{newtxtext,newtxmath}
\usepackage[varqu,scaled=0.92]{zi4}
\ifPDFTeX\else
\fi
\usepackage[american]{babel}
\usepackage{booktabs,array,tabularx}
\usepackage{graphicx}
\usepackage{newfloat}
\DeclareFloatingEnvironment[name=Listing,placement=!t]{listing}
\usepackage{microtype}
\usepackage{xcolor}
\usepackage{cite}
\usepackage[hidelinks]{hyperref}
\usepackage{xurl}
\hypersetup{pdftitle={Distinguishing Process Change, Instrument Fault and Data-Path Corruption in Streaming Telemetry: A Reference Implementation and Benchmark},pdfauthor={Oleksandr Klymenko}}
\usepackage{placeins}
\usepackage{dblfloatfix}
\begin{document}

\title{Distinguishing Process Change, Instrument Fault and Data-Path Corruption in Streaming Telemetry: A Reference Implementation and Benchmark}

\author{\IEEEauthorblockN{Oleksandr Klymenko}
\IEEEauthorblockA{Independent Researcher, Poland \\
ORCID 0009-0000-5924-3697 \\
alexanderklmn@gmail.com}}

\maketitle

\begin{abstract}
A reading departs from what a monitoring system expects for one of three reasons: the physical process changed, the instrument is failing, or the value was corrupted between the instrument and the software. The three call for different responses and look alike in a single sample. We state the three-cause problem, prove which pairs of causes one channel cannot separate, and propose a bank of online detectors --- a Kalman tracker with CUSUM and EWMA charts on its innovation, a Hampel median with generalized-ESD confirmation, a robust noise statistic with ADWIN, exact-value checks and an optional two-channel complementary filter --- combined by a deterministic, inspectable attribution rule. The detectors were merged into a public algorithm library. On 2,942 injected events of thirteen fault types the method detects 96.3\% (95\% interval 95.7--97.0) at 0.22 false alarms per hour and attributes 91.2\% (90.6--91.8) of detected events correctly from one channel and 99.0\% (98.6--99.4) with a second, rate channel; from one channel only 2 of 207 detected sensor drifts are attributed to the instrument, as the problem statement predicts. Injected into recorded industrial signals, the same catalogue is detected at 80.8\% and attributed correctly at 84.5\%, at 3.6 false alarms per hour. On the Skoltech Anomaly Benchmark the method detects 32 of 33 experiments with a median delay of 15~s, at 46 false alarms per hour against 11 for a rolling z-score.
\end{abstract}


\section{Introduction}

Monitoring systems for water treatment, pumping stations and production lines act on readings that arrive continuously from field instruments --- the multi-parameter water-quality measurements of an urban monitoring system~\cite{r1}, or the state of a machine tool archived to a cloud service through its controller~\cite{r2,r3} --- and the operator who receives them usually cannot audit the chain a reading traveled. When a reading departs from expectation, three answers are possible. \textbf{The process changed}: the measured quantity really moved, and the response is operational. \textbf{The instrument is failing}: it drifts, its noise grows, it sticks or saturates, its bias shifts, and the response is maintenance. \textbf{The data path corrupted the reading}: a transport, serialization or middleware fault replaced, re-sent, zeroed or lost the value, and the response is to repair the path and distrust the segment. Confusing them sends a technician to a working sensor, has operators react to a process change that did not happen, or discards a true event.

Change detection~\cite{r4,r5,r6,r7}, fault detection and isolation~\cite{r8,r9}, concept-drift detection~\cite{r10,r11}, sensor-fault diagnosis~\cite{r12}, data-quality work~\cite{r13} and time-series anomaly detection~\cite{r14,r15,r16,r17} each address one of the causes, or detect a departure without naming its cause. What is missing, in our reading, is a treatment of the three as one problem, posed where the monitoring software receives the stream and under its constraints: one pass, bounded memory per channel, no retraining, a decision within a bounded number of samples.

Our contributions are (i) a statement of the problem with two propositions that establish which pairs of causes one channel cannot separate (Section III); (ii) a detector bank and a deterministic attribution rule that returns a cause or \emph{undetermined} with its evidence (Section IV); (iii) a reference implementation whose detectors were reviewed and merged into a public library (Section V); and (iv) a measured account on 2,942 synthetic and 2,040 semi-synthetic injected events and a public industrial dataset, including where the method fails (Sections VI--VII). The detectors are classical by design: a system that runs for years on constrained hardware should rest on methods whose behavior is understood.

\section{Related Work}

Page's CUSUM~\cite{r4} and Roberts's EWMA chart~\cite{r5} are the standard online tests for a mean shift; Basseville and Nikiforov~\cite{r6} and Gustafsson~\cite{r7} give the theory of abrupt-change detection and its use in adaptive filtering. They detect that the distribution changed, not who changed it. Model-based fault detection and isolation~\cite{r8,r9} tests the residual of a plant model, canonically the Kalman innovation~\cite{r18}; we use the weakest useful model, a random-walk state, plus one physical fact, a bound on how fast the quantity can move. Sensor-fault surveys~\cite{r12} classify bias, drift, noise growth, failure and stuck-at, and a review of sensor-data quality~\cite{r15} catalogues the same errors from the data side; we take their signatures as the definition of the second and third causes. Surveys of anomaly detection in IoT time series~\cite{r14} and of outlier detection in time series~\cite{r16}, and a benchmark study of time-series outlier definitions~\cite{r17}, treat detection; attribution of a detected departure to its cause is outside their scope. Hampel's identifier~\cite{r19} and Rosner's generalized ESD test~\cite{r20} detect point anomalies, the signature of most data-path faults. Outlier-robust Kalman filters such as EMORF/S~\cite{r21} and its adaptive successor EMORF-II~\cite{r22} absorb outliers inside the filter by inferring, for each measurement, whether it is an outlier; we keep the outlier stage separate from the tracker, so that every rejected sample remains a visible event to which a cause can be attributed rather than a down-weighted measurement. Welford's recurrence~\cite{r23}, the P\textsuperscript{2} quantile estimator~\cite{r24} and ADWIN~\cite{r25} provide constant-memory statistics and adaptive windows.

\section{Problem Statement}

\subsection{Observation model}

\begin{table*}[!t]
\centering
\footnotesize
\setlength{\tabcolsep}{4pt}
\caption{The three causes and their signatures.}
\label{tab:1}
\begin{tabularx}{\textwidth}{>{\raggedright\arraybackslash}p{3.1cm}>{\raggedright\arraybackslash}X>{\raggedright\arraybackslash}X}
\toprule
\textbf{Cause} & \textbf{Term} & \textbf{Temporal signature} \\
\midrule
\textbf{Process change} & $x_t$ & level shift, ramp or regime change at a physically plausible rate; noise and bias unchanged \\
\textbf{Instrument fault} & $b_t$, $\sigma_t$, $h$ & bias step or drift; growth of $\sigma_t$; output stuck at the last value; saturation \\
\textbf{Data-path corruption} & $\mathcal{D}_t$ & isolated points or short bursts; segments exactly repeating earlier values; zeros, boundary, out-of-range or missing values \\
\bottomrule
\end{tabularx}
\end{table*}

Let $x_t$ be the quantity of interest and $y_t$ what the software receives:

\begin{equation}
y_t = \mathcal{D}_t\big(\, h(x_t) + b_t + n_t \,\big), \qquad n_t \sim \mathcal{N}(0, \sigma_t^2),
\label{eq:1}
\end{equation}

\noindent with $h$ the measurement function, $b_t$ the instrument bias, $n_t$ the measurement noise, and $\mathcal{D}_t$ the data-path operator, the identity when the path is healthy. In industrial architectures $\mathcal{D}_t$ is the chain from controller to supervisory system to cloud store~\cite{r2,r3}. The causes are departures in different terms (Table~\ref{tab:1}).

\subsection{What one channel cannot separate}

Let $h$ be the identity and $\mathcal{D}_t$ healthy, and let $u_t = \mathbb{1}[t \ge t_0]$.

\textbf{Proposition 1 (level shift and bias step).} For any $a$, the hypotheses $H_P$: $x'_t = x_t + a\,u_t$, $b'_t = b_t$ and $H_I$: $x'_t = x_t$, $b'_t = b_t + a\,u_t$ induce the same distribution of $(y_t)$. Hence no test on $(y_t)$ alone separates them better than chance.

\emph{Proof.} Under both hypotheses $y_t = x_t + b_t + a\,u_t + n_t$ with the same noise, so the laws coincide. $\square$

The separation therefore requires information outside $(y_t)$. One piece is physical: if the process cannot change faster than $\rho$ per sample, $|x_t - x_{t-1}| \le \rho$, a step completing in $m$ samples can be a process change only if $|a| \le m\rho$. A bias step larger than the process could produce in the observed time is identifiable; a smaller one is not (Section VII-D measures this boundary).

\textbf{Proposition 2 (drift and trend).} The same holds for $H_P$: $x'_t = x_t + \beta (t - t_0)^+$ and $H_I$: $b'_t = b_t + \beta (t - t_0)^+$ whenever the drifted process stays admissible, $|x_t - x_{t-1} + \beta| \le \rho$. If a second channel $r_t = x_t - x_{t-1} + \epsilon_t$ measures the rate of the process and not of the bias, the hypotheses become separable for every $\beta \ne 0$ as $t - t_0$ grows.

\emph{Proof.} The first part is Proposition 1 with $a\,u_t$ replaced by $\beta (t - t_0)^+$, which is admissible for the process under the stated condition. For the second, $\sum_{s \le t} r_s$ follows $x'_t$ under $H_P$ and $x_t$ under $H_I$, so $y_t - \sum_{s \le t} r_s$ has mean growing as $\beta (t - t_0)$ under $H_I$ only, while its noise, the sum of the $\epsilon_s$, grows only as $\sqrt{t - t_0}$. $\square$

Every other signature in Table~\ref{tab:1} --- noise growth, stuck-at, saturation, point anomalies, exact repeats, out-of-range values --- changes the law of $(y_t)$ in a way that no process path admissible under (1) with rate bound $\rho$ reproduces; we expect one channel to attribute these, and Section VII measures how well it does.

\subsection{Constraints}

One pass, state bounded independently of stream length, a decision within a bounded delay, no labeled training data, and thresholds calibrated on the channel's own data.

\section{Method}

\subsection{Calibration}

Each channel calibrates on its own data. In phase A (300 samples) the noise scale is estimated from first differences,

\begin{equation}
\hat\sigma = \frac{\operatorname{median}_t |y_t - y_{t-1}|}{0.6745\,\sqrt{2}},
\label{eq:2}
\end{equation}

\noindent with the median kept by P\textsuperscript{2}~\cite{r24}; for Gaussian noise $|y_t - y_{t-1}|$ has median $0.6745\sqrt{2}\,\sigma$, so (2) is consistent and insensitive to slow process movement. The channel is \emph{quantized} if more than 20\% of differences are zero or more than 5\% of values repeat an earlier value; $\hat\sigma$ is then the larger of (2) and the sample standard deviation, and the exact-value checks are disabled. In phase B (300 samples) the detectors run silently while P\textsuperscript{2} records the 99.9th percentiles $q_\eta$ and $q_d$ of the statistics (6) and (3); the thresholds are $1.25\,q$, never below the configured floors.

\subsection{Detectors}

\emph{Exact-value checks} act on the raw value: six identical consecutive values are \emph{stuck}; four consecutive values each exactly equal to an earlier value within 256 samples are a \emph{replayed segment}; a value equal to the running extremum over 16 samples and occurring three times in 64 is \emph{saturation}. A \emph{range guard} catches missing, out-of-range and declared boundary values before any filter.

\emph{Point anomalies.} With $m_t$ the median of the last 11 samples (Hampel filter~\cite{r19}), a sample is flagged when

\begin{equation}
d_t = \frac{|y_t - m_t|}{\hat\sigma} > \theta_d = \max(4.5,\; 1.25\,q_d),
\label{eq:3}
\end{equation}

\noindent and replaced by $m_t$ downstream. The scale is $\hat\sigma$, not the window's MAD, whose sampling error over 11 samples let 5--6$\sigma$ outliers through on development data. A flag counts as evidence only if GESD~\cite{r20} ($\alpha = 0.01$, last 31 samples, up to five outliers) confirms it.

\emph{Level changes.} A scalar Kalman filter~\cite{r18} with a random-walk state tracks the cleaned value $\tilde y_t$:

\begin{equation}
\begin{gathered}
\hat x_{t|t-1} = \hat x_{t-1},\quad P_{t|t-1} = P_{t-1} + Q,\quad S_t = P_{t|t-1} + R,\quad \\
z_t = \frac{\tilde y_t - \hat x_{t|t-1}}{\sqrt{S_t}},
\end{gathered}
\label{eq:4}
\end{equation}

\noindent with $Q = 0.002\,\hat\sigma^2$, $R = \hat\sigma^2$ and the usual update $K_t = P_{t|t-1}/S_t$, $\hat x_t = \hat x_{t|t-1} + K_t \sqrt{S_t}\, z_t$, $P_t = (1 - K_t) P_{t|t-1}$. Under a correct model $z_t$ is standard normal; the two-sided CUSUM~\cite{r4} and the EWMA chart~\cite{r5} run on it:

\begin{equation}
\begin{gathered}
g^{\pm}_t = \max\!\big(0,\; g^{\pm}_{t-1} \pm z_t - k\big),\ \text{alarm if } g^{\pm}_t > H; \\
e_t = \lambda z_t + (1 - \lambda) e_{t-1},\ \text{alarm if } |e_t| > L\sqrt{\tfrac{\lambda}{2-\lambda}\big(1 - (1-\lambda)^{2\ell}\big)},
\end{gathered}
\label{eq:5}
\end{equation}

\noindent with $k = 0.5$, $H = 8$, $\lambda = 0.1$, $L = 4$, $\ell$ the samples since the last reset, and both statistics reset on alarm and after each verdict.

\emph{Noise growth.} The noise statistic is the current noise level relative to calibration,

\begin{equation}
\begin{gathered}
\eta_t = \frac{\operatorname{median}_{0 \le i < 40} |y_{t-i} - y_{t-i-1}|}{0.6745\,\sqrt{2}\,\hat\sigma}, \\
\text{noise if } \eta_t > \theta_\eta = \max(1.8,\; 1.25\,q_\eta).
\end{gathered}
\label{eq:6}
\end{equation}

A smooth process movement changes $|y_t - y_{t-1}|$ by at most $\rho$, one outlier does not move a median of 40, and a noisier instrument scales $\eta_t$ by its noise factor. ADWIN~\cite{r25} watches the mean of the clipped $|y_t - y_{t-1}|/\hat\sigma$. A first design used the windowed variance of $z_t$ and sent process ramps to the instrument, because a ramp biases the innovation.

\subsection{Second channel}

Where a second channel measures the rate $r_t$ of the same quantity, a complementary filter fuses the integrated rate with the level,

\begin{equation}
\check x_t = c\,(\check x_{t-1} + r_t) + (1 - c)\,\tilde y_t, \qquad \delta_t = \operatorname{median}_{0 \le i < 15} \frac{\tilde y_{t-i} - \check x_{t-i}}{\hat\sigma},
\label{eq:7}
\end{equation}

\noindent with $c = 0.995$ (time constant about 200 samples). By Proposition 2, $\delta_t$ stays near zero when the process moves and follows the bias when the bias steps or drifts.

\subsection{Attribution rule}

The first detector to fire, at $t_0$, opens a decision window of $\Delta = 40$ samples. At its end, $t_1 = t_0 + \Delta$, the rule reads three features of the cleaned series:

\begin{equation}
\begin{gathered}
s = \frac{\big|\operatorname{med}_9(\tilde y)_{t_1} - \operatorname{med}_9(\tilde y)_{t_0^-}\big|}{\hat\sigma}, \\
\gamma = \frac{\hat b_{60}(t_1)}{\hat\sigma}, \\
j = \max_{t_0 \le t \le t_1} \frac{\big|\operatorname{med}_5(\tilde y)_t - \operatorname{med}_5(\tilde y)_{t-5}\big|}{\hat\sigma},
\end{gathered}
\label{eq:8}
\end{equation}

\begin{table*}[!t]
\centering
\footnotesize
\setlength{\tabcolsep}{4pt}
\caption{Attribution rule; parameters are the frozen configuration.}
\label{tab:2}
\begin{tabularx}{\textwidth}{>{\raggedright\arraybackslash}p{3.1cm}>{\raggedright\arraybackslash}X>{\raggedright\arraybackslash}X}
\toprule
\textbf{\#} & \textbf{Evidence in the window} & \textbf{Attribution} \\
\midrule
1 & range guard (missing, out of range, boundary) & \textbf{data-path} \\
2 & replayed segment & \textbf{data-path} \\
3 & saturation & \textbf{instrument} \\
4 & stuck at the last value & \textbf{instrument} \\
5 & (6) above threshold, still at decision; $s < 2\max(1, \eta_{t_1})$ & \textbf{instrument} \\
6 & 1--4 GESD-confirmed outliers; $s < 2$ & \textbf{data-path} \\
7 & level change ($s \ge 2$, or (5) alarm while moving), $j > j^*$ & \textbf{instrument} \\
8 & level change; $\lvert\delta_{t_1}\rvert > 2.5$ & \textbf{instrument} \\
9 & level change; $\lvert\delta_{t_1}\rvert \le 2.5$, settled or horizon reached & \textbf{process} \\
10 & level change, one channel, settled & \textbf{process} \\
11 & level change, one channel, moving & defer, up to 240 samples \\
12 & level change, one channel, moving at 240 & \textbf{undetermined} \\
13 & $\geq$ 5 GESD-confirmed outliers, no level change & \textbf{data-path} \\
14 & none & window dismissed \\
\bottomrule
\end{tabularx}
\end{table*}

\noindent where $\operatorname{med}_k(\tilde y)_t$ is the median of the $k$ cleaned samples ending at $t$ and $t_0^-$ is 4 samples before the trigger (or the baseline carried over from a continued window): the net level change, the least-squares slope over the last 60 samples, and the largest change between 5-sample medians five samples apart, taken from $t_0$ until the verdict, including any deferral. The level is \emph{moving} if $|\gamma| > 0.03$. A process limited to $\rho$ per sample moves the 5-sample median by at most $5\rho$ in 5 samples; the median of 5 Gaussian samples has standard deviation at most about $\sigma\sqrt{\pi/10} \approx 0.56\,\sigma$, so the difference of two has about $0.79\,\sigma$, and three of those give a per-comparison margin:

\begin{equation}
j > j^* = \frac{5\rho}{\hat\sigma} + 2.4 \quad \Longrightarrow \quad \text{faster than the process can move.}
\label{eq:9}
\end{equation}

The first matching row of Table~\ref{tab:2} decides.

After a verdict, the kinds of evidence behind it are latched for a 40-sample hold-off that they extend while they keep firing, so one persistent condition yields one verdict; other evidence opens a new window at once. A dismissed window latches nothing, and a change alarm within 240 samples of it continues it with its baseline, so that a change too slow to reach $2\hat\sigma$ in $\Delta$ samples accumulates instead of being dismissed repeatedly. Every verdict carries the evidence of the rows that matched, so an operator can see why the cause was assigned.

\section{Reference Implementation}

\begin{table*}[!t]
\centering
\footnotesize
\setlength{\tabcolsep}{4pt}
\caption{Library components used in the evaluated bank. Three further components from the same series --- extended and unscented Kalman filters and dynamic time warping (TheAlgorithms/Java\#7615, TheAlgorithms/Java\#7622, TheAlgorithms/Java\#7619) --- are not used here.}
\label{tab:3}
\begin{tabularx}{\textwidth}{>{\raggedright\arraybackslash}p{3.1cm}>{\raggedright\arraybackslash}X>{\raggedright\arraybackslash}X}
\toprule
\textbf{Component} & \textbf{Role} & \textbf{Pull request} \\
\midrule
\texttt{KalmanFilter} & tracker (4) & TheAlgorithms/\allowbreak{}Java\#7592 \\
\texttt{CusumDetector}, \texttt{EwmaChangeDetector} & charts (5) & TheAlgorithms/\allowbreak{}Java\#7593, TheAlgorithms/\allowbreak{}Java\#7600 \\
\texttt{HampelFilter}, \texttt{GeneralizedEsd} & running median (3); confirmation & TheAlgorithms/\allowbreak{}Java\#7598, TheAlgorithms/\allowbreak{}Java\#7612 \\
\texttt{WelfordAlgorithm}, \texttt{P2QuantileEstimator} & calibration (2), thresholds & TheAlgorithms/\allowbreak{}Java\#7602, TheAlgorithms/\allowbreak{}Java\#7603 \\
\texttt{Adwin} & sustained noise change & TheAlgorithms/\allowbreak{}Java\#7606 \\
\texttt{ComplementaryFilter} & second channel (7) & TheAlgorithms/\allowbreak{}Java\#7614 \\
\texttt{SlidingWindowAggregator} & running extremum & TheAlgorithms/\allowbreak{}Java\#7591 \\
\bottomrule
\end{tabularx}
\end{table*}

The detectors were written in Java without external dependencies and, between 6 and 29 September 2026, submitted to TheAlgorithms/Java, a public algorithm library with 66,339 GitHub stars and 21,276 forks as of 2 October 2026; each was reviewed, approved and merged by a project maintainer other than the author (Table~\ref{tab:3}). The benchmark vendors them unmodified from one pinned commit, \texttt{3f067087f217afc6c16f0d0ca90d194bd7a55726}.

The harness --- process models, injectors, scoring, baselines, a SKAB reader and SVG plots --- is released under the MIT license at \url{https://github.com/alxkm/telemetry-triage}; one entry point reproduces every number and figure. The configuration was frozen before the test runs, and the repository lists each of the nine changes made on development data. A channel's state is bounded independently of stream length: 1,296 doubles in configuration-sized arrays, 32 boxed values in the extremum aggregators, and ADWIN buckets growing logarithmically (at most 31 over $10^6$ samples). One update takes a median of 5.8 $\mu$s on one core (OpenJDK 21, wall-clock over $10^6$ samples after warm-up, not JMH).

\section{Benchmark}

\begin{table*}[!t]
\centering
\footnotesize
\setlength{\tabcolsep}{4pt}
\caption{Injection catalogue. Types are drawn in shuffled rounds and separated by at least 420 samples. Development seeds 1000--1019 fixed the configuration; results use test seeds 2000--2049 per model (2,942 events), run once after the freeze. For a given seed each process model draws its own random stream, so the two models share neither noise nor fault schedule.}
\label{tab:4}
\begin{tabularx}{\textwidth}{>{\raggedright\arraybackslash}p{3.1cm}>{\raggedright\arraybackslash}X>{\raggedright\arraybackslash}X}
\toprule
\textbf{Injection} & \textbf{Cause} & \textbf{Parameters} \\
\midrule
level shift; ramp; regime change & process & $\pm$4--10$\sigma$; $\pm$0.02--0.08$\sigma$/sample for 100--300; $\pm$3--6$\sigma$ for 300--500 with a 1--2$\sigma$ set-point oscillation of period 100--200 \\
bias step; linear drift & instrument & $\pm$4--10$\sigma$; $\pm$0.02--0.08$\sigma$/sample for 100--300, both persisting \\
noise growth; stuck; saturation & instrument & $\times$2.5--5 for 200--400; 30--200 samples; for 200--400, ceiling 0.3--0.8$\sigma$ above the noise-free level plus bias \\
spike; substitution burst; replay & data-path & 1 sample $\pm$6--15$\sigma$; 3 samples $\pm$6--15$\sigma$; 10--40 samples re-sent from 50--200 earlier \\
zero; dropout & data-path & 1--3 samples; 5--20 missing samples \\
\bottomrule
\end{tabularx}
\end{table*}

\textbf{Synthetic streams.} Two process models generate $x_t$ in units of the instrument noise $\sigma$: a \emph{tank} whose level moves toward its set-point at a pump-limited rate ($\rho = 0.15\sigma$ per sample; $\sigma = 0.5$ cm), and a \emph{heated vessel} with a first-order response ($\tau = 60$ samples; $\sigma = 0.1$ \textdegree{}C; $\rho = 0.2\sigma$). Both carry an Ornstein--Uhlenbeck wander (stationary standard deviation $0.5\sigma$, correlation time 500 samples) and Gaussian noise. A second channel measures the true rate of change with noise $0.05\sigma$ and is unaffected by instrument and data-path faults. Runs have 20,000 samples; one sample counts as one second.

\textbf{Semi-synthetic streams.} To test the method on real noise and real background dynamics, the same catalogue and schedule are injected into each of the eight recorded channels of SKAB's anomaly-free file~\cite{r26} (9,405 rows at 1 Hz): process changes as set-point moves limited to $0.15\hat\sigma$ per sample added to the recording, declared as the plausible rate; instrument and data-path faults as in Table~\ref{tab:4}, with amplitudes in units of the $\hat\sigma$ the method itself estimates in calibration. Seeds 3000--3019 per channel give 2,040 events. There is no rate channel. This file was used during development to fix two calibration rules (the data-driven outlier threshold in (3) and the quantization test); no attribution threshold was tuned on it.

\textbf{SKAB.} The Skoltech Anomaly Benchmark~\cite{r26} records eight sensors of a water-circulation rig at 1 Hz with anomalies induced by valve and pump manipulation. It labels anomalies, not causes, so it serves detection delay and false alarms; attribution is reported descriptively. One of 34 labeled files (\texttt{other/2.csv}) starts its anomaly before calibration ends and is excluded. Calibration is shortened to 200 + 200 rows, since anomalies begin near row 570. A file is detected when any channel triggers on a labeled row.

\textbf{Metrics.} An event is matched by the first decision whose trigger falls in [onset, end + 120 samples]; delay is trigger minus onset. Further decisions inside a matched window are counted separately and are neither matches nor false alarms. Attribution is a 3$\times$4 confusion matrix (three causes plus \emph{undetermined}); precision is over matched decisions, recall over all events. False alarms are decisions triggered outside all event windows, per monitored hour. Intervals are 95\% percentile bootstrap intervals over runs (2,000 resamples of the 100 test runs).

\textbf{Baselines.} A rolling z-score (300-sample window, $|z| > 4$, hold-off 40) and a restarted CUSUM on the raw value ($k = 0.5$, $H = 8$; reference relearned from 40 samples after each alarm). Neither attributes.

\section{Results}

\begin{table*}[!t]
\centering
\footnotesize
\setlength{\tabcolsep}{4pt}
\caption{Both models pooled, test seeds. \emph{Undetermined} is a share of detected events and counts as a miss in recall. The low one-channel precision for \emph{process} is Proposition 2: 176 of the 178 instrument events called \emph{process} are sensor drifts. Over all causes, one channel: detected 96.3\% [95.7, 97.0], correct of detected 91.2\% [90.6, 91.8], undetermined 2.1\% [1.6, 2.7]; two channels: correct of detected 99.0\% [98.6, 99.4]; false alarms 0.216 per hour [0.160, 0.278] (73 in 338.1 h).}
\label{tab:5}
\begin{tabularx}{\textwidth}{>{\raggedright\arraybackslash}p{3.1cm}>{\centering\arraybackslash}X>{\centering\arraybackslash}X>{\centering\arraybackslash}X>{\centering\arraybackslash}X>{\centering\arraybackslash}X}
\toprule
\textbf{True cause} & \textbf{Events} & \textbf{Detected (one channel)} & \textbf{One channel: P / R / F1} & \textbf{Undet.} & \textbf{Two channels: P / R / F1} \\
\midrule
process & 664 & 89.2\% & 75.4 / 83.0 / 79.0\% & 5.2\% & 97.3 / 87.5 / 92.1\% \\
instrument & 1,129 & 97.5\% & 99.0 / 79.2 / 88.0\% & 2.5\% & 99.1 / 96.7 / 97.9\% \\
data-path & 1,149 & 99.3\% & 99.8 / 99.1 / 99.5\% & 0.0\% & 99.8 / 99.1 / 99.5\% \\
\bottomrule
\end{tabularx}
\end{table*}

\begin{figure*}[!p]
\centering
\includegraphics[width=\textwidth,height=0.86\textheight,keepaspectratio]{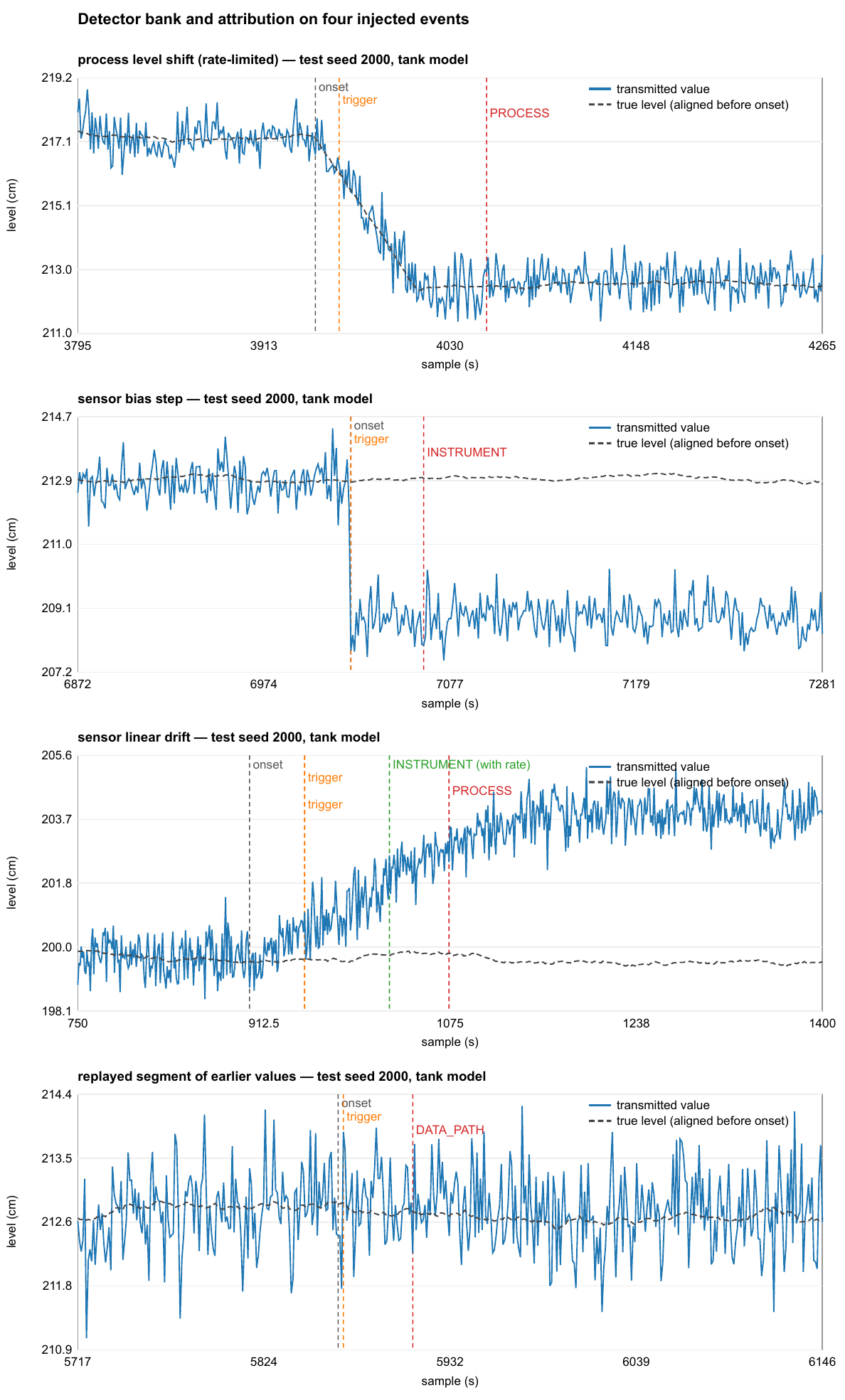}
\caption{Four injected events on test seed 2000, tank model: transmitted value, true level (aligned to the pre-onset offset), onset, first trigger and verdict. In the drift panel the one-channel verdict is \emph{process}; with the rate channel it is \emph{instrument}.}
\label{fig:1}
\end{figure*}

\subsection{Synthetic streams}

Table~\ref{tab:5} gives detection and attribution by true cause. Median detection delay is 3~s (P90 33~s): 0~s for data-path, 4~s for instrument and 20~s for process events, which develop at a bounded rate; the verdict follows after the 40-sample window. Every point-like and sequence-level signature (noise growth, stuck, saturation and the five data-path types) is attributed at 99.1--100\%. The one systematic failure is the predicted one: from one channel only 2 of the 207 detected drifts are attributed to the instrument --- 176 are called \emph{process}, 28 \emph{undetermined}, 1 \emph{data-path} --- and with the rate channel 198 of 213 detected drifts (93.0\%) are. Fig.~\ref{fig:1} shows four events with their triggers and verdicts.

\textbf{Missed events.} Of the 108 events missed from one channel, 47 were \emph{absorbed}: a decision window opened shortly before the onset by an earlier trigger was still open at the onset, so the event was decided inside that window and is scored as a miss and a false alarm. The other 61 were \emph{silent}: no detector fired within the event window. Silent misses are slow process changes and drifts --- 20 regime changes, 20 ramps and 15 drifts --- whose rate keeps the innovation within the charts' limits, and 6 level shifts; no point-like or sequence-level fault was missed silently.

\subsection{Semi-synthetic streams}

\begin{table}[!t]
\centering
\footnotesize
\setlength{\tabcolsep}{4pt}
\caption{The catalogue injected into recorded SKAB channels, level channel only. \textdagger{} Quantized channel (Section IV-A). Pooled P / R / F1: process 72.2 / 59.3 / 65.1\%, instrument 82.0 / 62.9 / 71.2\%, data-path 98.6 / 79.1 / 87.8\%.}
\label{tab:6}
\begin{tabularx}{\columnwidth}{>{\raggedright\arraybackslash}X>{\centering\arraybackslash}X>{\centering\arraybackslash}X>{\centering\arraybackslash}X}
\toprule
\textbf{Channel} & \textbf{Detected} & \textbf{Correct of detected} & \textbf{False alarms per hour} \\
\midrule
Accelerometer 1 \textperiodcentered{} 2 & 82.7\% \textperiodcentered{} 93.7\% & 83.9\% \textperiodcentered{} 91.6\% & 9.95 \textperiodcentered{} 1.93 \\
Current \textperiodcentered{} Voltage & 80.4\% \textperiodcentered{} 96.1\% & 75.1\% \textperiodcentered{} 91.4\% & 10.04 \textperiodcentered{} 0.06 \\
Pressure \textperiodcentered{} Volume flow\textdagger{} & 76.1\% \textperiodcentered{} 74.1\% & 86.6\% \textperiodcentered{} 81.0\% & 0.85 \textperiodcentered{} 3.19 \\
Temperature \textperiodcentered{} Thermocouple\textdagger{} & 69.8\% \textperiodcentered{} 73.7\% & 74.7\% \textperiodcentered{} 87.8\% & 2.27 \textperiodcentered{} 0.51 \\
\textbf{all eight} (2,040 events) & \textbf{80.8\%} & \textbf{84.5\%} & \textbf{3.60} \\
\bottomrule
\end{tabularx}
\end{table}

\begin{table*}[!t]
\centering
\footnotesize
\setlength{\tabcolsep}{4pt}
\caption{Detection against baselines (no baseline attributes). On synthetic data the z-score detects 1 of 227 replayed segments and almost no dropout, which it skips; the restarted CUSUM detects every process change but raises 9.6 false alarms per hour. SKAB: 2.00 h of unlabeled rows in labeled files, 2.50 h in the anomaly-free file.}
\label{tab:7}
\begin{tabularx}{\textwidth}{>{\raggedright\arraybackslash}p{3.1cm}>{\centering\arraybackslash}X>{\centering\arraybackslash}X>{\centering\arraybackslash}X>{\centering\arraybackslash}X>{\centering\arraybackslash}X>{\centering\arraybackslash}X}
\toprule
\textbf{} & \textbf{Synthetic: detected} & \textbf{Synthetic: delay med/P90 (s)} & \textbf{Synthetic: FA/h} & \textbf{SKAB: detected (of 33)} & \textbf{SKAB: delay med/P90 (s)} & \textbf{SKAB: FA/h (labeled / anomaly-free)} \\
\midrule
this method & 96.3\% & 3 / 33 & 0.22 & 32 & 15 / 54 & 46.0 / 34.0 \\
rolling z-score & 62.4\% & 0 / 82 & 0.40 & 33 & 35 / 66 & 11.0 / 12.0 \\
restarted CUSUM & 82.0\% & 13 / 74 & 9.61 & 33 & 5 / 22 & 179.5 / 173.1 \\
\bottomrule
\end{tabularx}
\end{table*}

On real noise (Table~\ref{tab:6}) the method detects 80.8\% of injected events and attributes 84.5\% of them correctly, against 96.3\% and 91.2\% on synthetic streams; false alarms rise from 0.22 to 3.6 per hour. Two effects are visible by fault type (repository, experiment E9). On the two quantized channels the exact-value checks are disabled by design (Section IV-A), so stuck values and replayed segments there are found only when other detectors fire --- 62.5\% and 63.8\% of them are detected over all channels, against 99.6\% and 99.1\% on synthetic streams. And the recorded background at times moves faster than the declared plausible rate, which costs process attribution: 24 of the 112 detected level shifts were called \emph{instrument} by the jump test (9). Data-path attribution stays high (precision 98.6\%). The false alarms concentrate on the two channels with the most background movement (accelerometer 1 and current, about 10 per hour each); over all channels, 300 of the 912 false alarms are attributed to the process and 490 to the instrument.

\subsection{Baselines and SKAB}

\begin{figure*}[!t]
\centering
\includegraphics[width=0.86\textwidth]{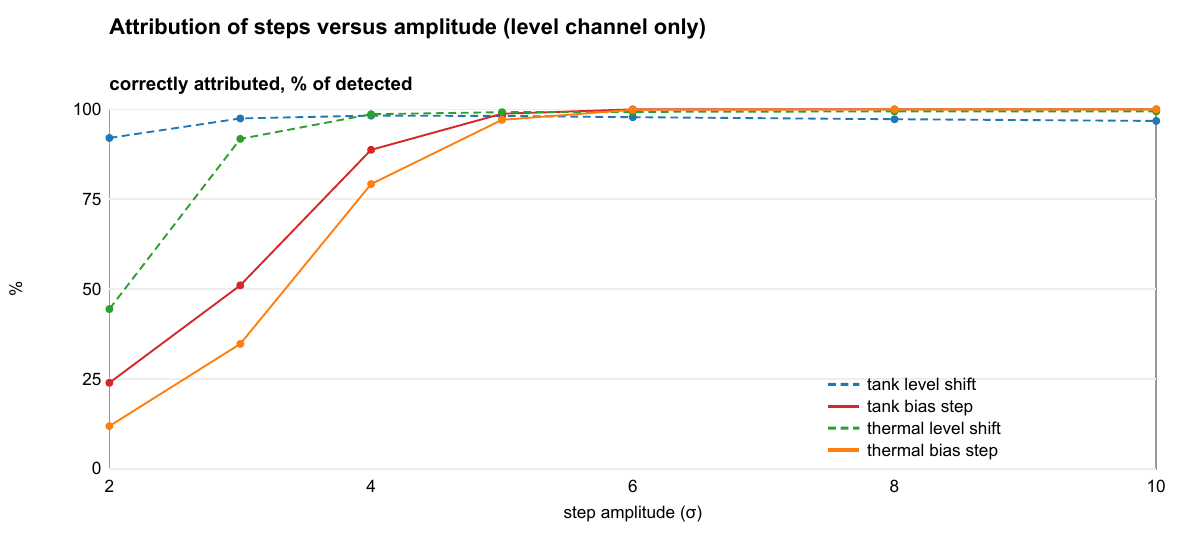}
\caption{Share of detected steps attributed correctly versus amplitude, one channel; 50 test seeds per model and amplitude. The vessel's level-shift value at $2\sigma$ rests on 9 detected events.}
\label{fig:2}
\end{figure*}

On SKAB (Table~\ref{tab:7}) the method detects 32 of 33 experiments in less than half the z-score's median delay (15~s against 35~s) but raises four times its false alarms. Of its 92 false alarms on labeled files, 80 are attributed to the \emph{process}: level changes of the rig that the dataset does not label. The first verdict in each detected anomaly was \emph{process} in 21 files, \emph{data-path} in 6, \emph{instrument} in 3 and \emph{undetermined} in 2; SKAB carries no cause labels, so this is not scored.

\subsection{Bias step versus process shift on one channel}

Fig.~\ref{fig:2} varies the step amplitude. From $3\sigma$ on, process shifts are attributed to the process at 91.8--99.4\%; at $2\sigma$ the tank's are at 92.0\%, while the vessel, whose first-order response spreads a $2\sigma$ shift over $\tau = 60$ samples, detects only 9 of 897 such shifts. Bias steps are attributed to the instrument at 24\% (tank) and 12\% (vessel) at $2\sigma$, 89\% and 79\% at $4\sigma$, and 97.1--100\% from $5\sigma$. The crossover sits at the bound (9), $j^* = 3.15$ and $3.40$: below it the step cannot be told from a rate-limited process move at this noise level, which is the identifiability limit discussed after Proposition 1.

\textbf{Sensitivity and ablations.} With $\Delta$ = 20 or 80 instead of 40, correct attribution is 88.7\% or 91.0\% (versus 91.2\%) at median verdict latencies of 23 and 83~s (versus 43~s). Without the plausible-rate test, bias-step attribution falls from 99.1\% to 0\%; without deferral, level-shift attribution falls from 99.0\% to 0\% and 28.5\% of detected events become \emph{undetermined}.

\section{Discussion and Limitations}

\textbf{What one channel cannot do.} Propositions 1 and 2 are borne out: sensor drift is not separated from a process trend (2 of 207 attributed to the instrument), nor a bias step below the bound (9) from a process shift. A method that names a confident cause there is guessing; on synthetic streams a second channel raises correct attribution from 91.2\% to 99.0\%.

\textbf{Real data.} On recorded signals the method keeps its data-path attribution but loses detection on quantized channels and process attribution where the background is fast; on SKAB it is faster and noisier than a rolling z-score. Most of its false alarms arrive labeled as process movements, which makes them triageable, but they remain alarms; a deployment would require a minimum event duration per channel. We did not tune on SKAB's labeled files.

\textbf{The model of expectation.} The tracker (4) is the simplest model of what a reading should be. A learned forecaster --- for instance the neural models for time-series forecasting examined in~\cite{r27} --- would sharpen the innovation $z_t$ and with it every chart in (5), at the cost of training and of the inspectability the rule is built for.

\textbf{Threats to validity.} The injections encode our assumptions, and the benchmark was designed by the method's author; separating development and test seeds limits but does not remove that bias, and the semi-synthetic base file was seen in development. Datasets with cause labels are scarce; SKAB supports detection, not attribution. A trigger a few samples before an onset is scored as a miss plus a false alarm (47 of 108 misses). Timing is from one machine.

\section{Conclusion}

We stated anomaly attribution in streaming telemetry as a three-cause problem, proved which causes one channel cannot separate, and measured an inspectable rule over classical online detectors: 91.2\% of detected events attributed correctly from one channel and 99.0\% from two, at 0.22 false alarms per hour on synthetic streams, and 84.5\% on recorded signals with injected faults --- failing where the propositions say one channel must. On real data it detects quickly but alarms too often for unattended use. Next: per-channel duration requirements, multi-channel plants with a physical model, and labeled industrial datasets as they become available.

\section*{Code and Data Availability}

The implementation, the benchmark harness, the frozen configuration and every result file are available at \url{https://github.com/alxkm/telemetry-triage} (commit \texttt{3246e44}); one command reproduces every number, table and figure of this paper. SKAB~\cite{r26} is not redistributed (GPL-3.0); a script in the repository downloads it and verifies its checksums.

\FloatBarrier
\makeatletter\ifx\@dbldeferlist\@empty\else\clearpage\fi\makeatother

\end{document}